\documentclass[10pt,b5paper,twoside,onecolumn]{scrartcl}
\usepackage[cp1250]{inputenc}
\usepackage{color,fancyvrb}
\usepackage{graphicx}
\usepackage{hyperref}
\usepackage{fancyhdr}
\usepackage{setspace}
\usepackage{float}
\usepackage{pdfpages}
\usepackage{verbatim}
\usepackage{lmodern} 
\usepackage{amsthm}
\usepackage{amssymb}
\usepackage{multicol}
\usepackage{mathpazo}
\usepackage{booktabs}
\usepackage[T1]{fontenc}
\usepackage[titles]{tocloft}
\begin{document}
	
\pagestyle{fancyplain}
\fancyhf{}
\fancyhead[LE]{\textit{An Order-based Algorithm for Minimum Dominating Set}}
\fancyhead[RO]{\textit{D. Chalupa}}
\fancyfoot[C]{\thepage}
\fancypagestyle{plain}
{
	\fancyhf{} 
	\renewcommand{\headrulewidth}{0pt} 
	\renewcommand{\footrulewidth}{0pt}
}

\thispagestyle{empty}
	
\begin{center}\textbf{\LARGE\sffamily\noindent
An Order-based Algorithm for Minimum Dominating Set with Application in Graph Mining
}\end{center}

\begin{center}{\large\sffamily\noindent David Chalupa$^{a,b}$}\end{center}
\begin{center}
{
\noindent
$^a$~Operations Research Group\\
Department of Materials and Production\\
Aalborg University\\
Fibigerstr\ae de 16, Aalborg 9220, Denmark\\
$^b$~Computer Science\\
School of Engineering and Computer Science\\
University of Hull\\
Cottingham Road\\
Hull HU6 7RX, United Kingdom\\
Email: \texttt{dc@m-tech.aau.dk}
} \vspace{15pt}\\
\textit{Acknowledgement:} This is a preprint version of the article. Reference to the definitive version is the following:\\
David Chalupa, An order-based algorithm for minimum dominating set with application in graph mining, In Information Sciences, Volume 426, 2018, Pages 101-116, ISSN 0020-0255.\\
DOI:~~\url{https://doi.org/10.1016/j.ins.2017.10.033}.
\end{center}

\vspace{15pt}
	
\paragraph{Abstract.} Dominating set is a set of vertices of a graph such that all other vertices have a neighbour in the dominating set.
We propose a new order-based randomised local search (RLS$_o$) algorithm to solve minimum dominating set problem in large graphs.
Experimental evaluation is presented for multiple types of problem instances.
These instances include unit disk graphs, which represent a model of wireless networks, random scale-free networks, as well as samples from two social networks and real-world graphs studied in network science.
Our experiments indicate that RLS$_o$ performs better than both a classical greedy approximation algorithm and two metaheuristic algorithms based on ant colony optimisation and local search.
The order-based algorithm is able to find small dominating sets for graphs with tens of thousands of vertices.
In addition, we propose a multi-start variant of RLS$_o$ that is suitable for solving the minimum weight dominating set problem. 
The application of RLS$_o$ in graph mining is also briefly demonstrated.


\section{Introduction}

Dominating set of a graph is a set of its vertices such that each vertex is in the dominating set or has a neighbour in the dominating set. Dominating sets and their variants have applications in several diverse areas, including routing in wireless ad-hoc networks \cite{dominatingwirelesslocal}, multi-document summarisation \cite{docsummarydominating} or modelling and studying of positive influence in social networks \cite{pidsapprox}. The problem of finding the minimum dominating set (MDS) is widely known to be NP-hard \cite{michael1979computers,parekh1991analysis,haynes1997domination}.

Let $G = [V,E]$ be an undirected graph and let $S \subseteq V$. Then, $S$ is a \textit{dominating set} if each vertex $v \in V$ is in $S$ or is adjacent to a vertex in $S$. Dominating set with the lowest cardinality is called \textit{minimum dominating set}, its cardinality is called \textit{domination number} and is denoted by $\gamma$.

Figure 1 presents an illustration of the dominating set for a sample of a social graph from Google+ with 200 vertices. Both drawings in the figure represent the same data. On the left hand side, the vertex with maximum degree is placed in the middle and other vertices are arranged into levels, based on their distance from the central vertex. On the right hand side, a dominance drawing of the network is presented, in which the dominating set is used to visually organise the network \cite{graphdrawingvisualization}. Vertices of the dominating set are highlighted in red and other vertices are grouped into clusters, effectively using the fact that each of these vertices has a neighbour in the dominating set. Dominating set can be used as a set of hubs of the network to form fine-grained clusters, since all other vertices have a neighbour in the dominating set.
	
\begin{figure}
\begin{center}
\includegraphics[scale=0.85, bb=0 0 189 189]{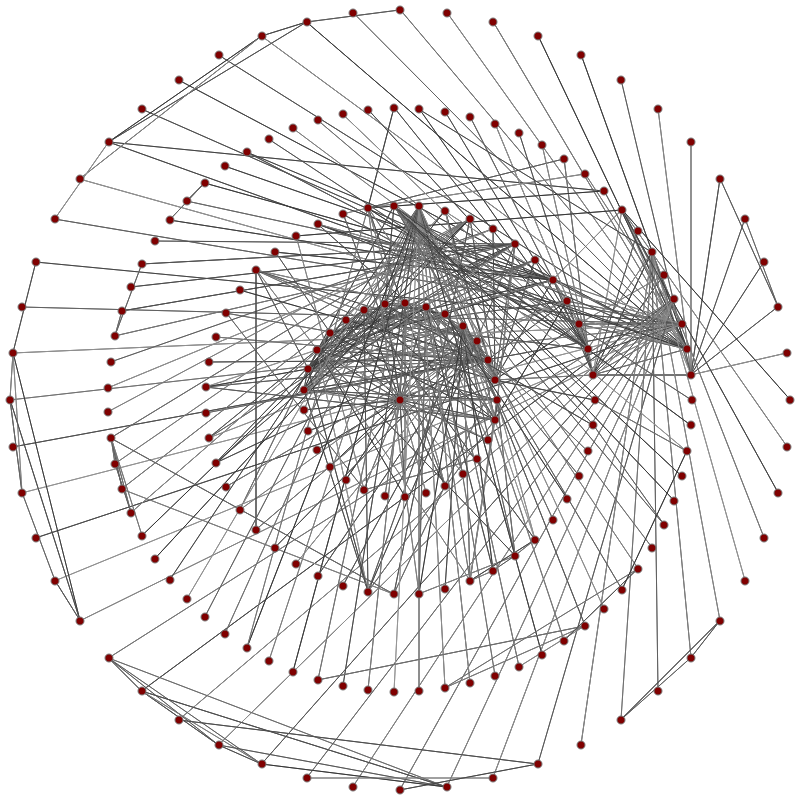}
\includegraphics[scale=0.89, bb=0 0 189 189]{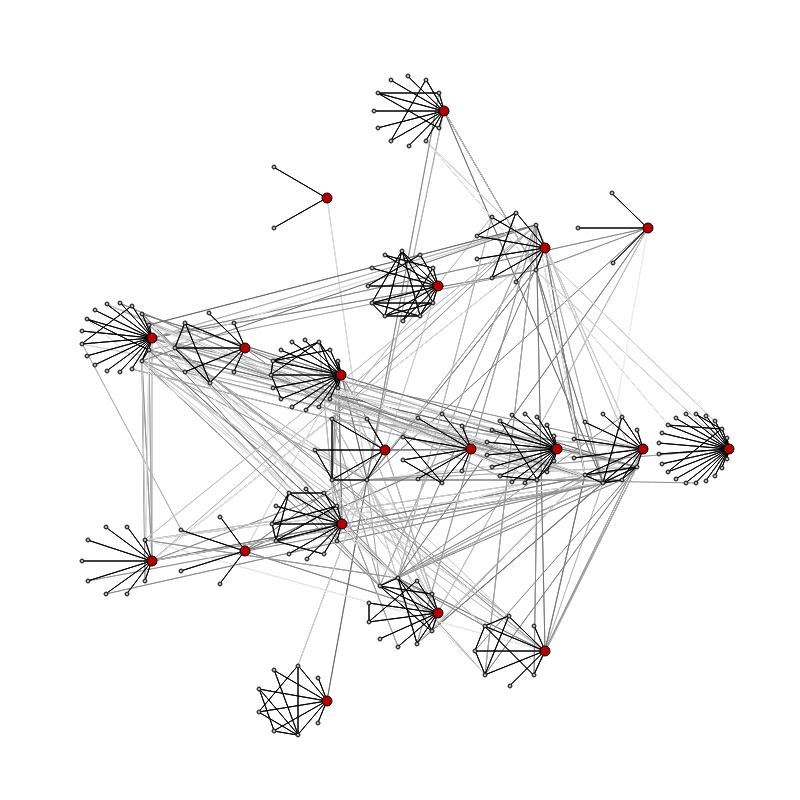}
\end{center}
\caption{Two visualisations of a sample social graph from Google+ with 200 vertices. On the left hand side, the vertex with maximum degree is taken as the central vertex and other vertices are visualised in a radial drawing, based on distance from the centre. On the right hand side, vertices are partitioned around vertices in the dominating set.}
\end{figure}

There is a large body of literature on dominating sets and their variations. The main focus in MDS literature is on theoretical aspects and applications in wireless ad-hoc networks. NP-hardness and approximation characteristics of the problem suggest that works aimed at design of efficient heuristics and their scalability are of an interest. However, solving techniques for MDS are also of an increasing interest for large graphs due to their applications in social and information networks.

The current core of experimental literature is focused mainly on benchmarking and applications of algorithms for weighted dominating sets \cite{jovanovic2010ant,dominatingweight,bouamama2016hybrid,wang2017local,lin2016effective} and connected dominating set problems \cite{dominatingwirelesslocal,acodominating}. Design of scalable algorithms for large-scale instances of MDS and their real-world applications in social and information networks seem to be the areas to explore further. In this paper, we propose a new heuristic algorithm for MDS, which is highly scalable to large graphs and tends to perform better than a greedy approximation algorithm, as well as algorithms based on constructive approaches and ant colony optimisation with local search.

Our \textit{order-based randomised local search} (RLS$_o$) algorithm tackles MDS indirectly by using a representation based on permutations of vertices, which are transformed into dominating sets using a greedy algorithm. The permutation of vertices is then optimised by repeated using of specific randomised $jump$ moves.

Experiments were carried out using a wide collection of both synthetic and real-world graphs. We compare RLS$_o$ to a classical greedy approximation algorithm, as well as a more recently proposed ant colony optimisation algorithm hybridised with local search (ACO-LS) \cite{twoheurdomset} and its extension with preprocessing (ACO-PP-LS) \cite{dominatingweight}. ACO-PP-LS was originally designed for the minimum weight dominating set problem.

We first present results of RLS$_o$ for unit disk graphs, since these graphs were used for evaluation in previous studies as models of wireless networks. Additionally, we provide results for scale-free networks generated by Barab\'{a}si-Albert (BA) model. We generally obtain that RLS$_o$ performs better than the greedy approximation algorithm, ACO-LS and ACO-PP-LS.

Results are also provided for an extensive collection of real-world graphs studied in network science, instances from DIMACS series, as well as samples of anonymised publicly available data from social networks \textit{Google+} and \textit{Pokec}. These results also confirm that RLS$_o$ provides results of better quality than the greedy approximation algorithm, ACO-LS and ACO-PP-LS, while maintaining solid scalability for large graphs. In addition, the solutions found by RLS$_o$ tend to be close to lower bounds, which have been computed as solutions to the linear programming relaxation of MDS. This relaxation represents the linear programming problem obtained from the formulation of MDS by assuming that decision variables can be any real values between $0$ and $1$, instead of binary variables.

RLS$_o$ is also extended to a multi-start algorithm MSRLS$_o$ to solve the minimum weight dominating set (MWDS) problem.  Finally, an application of RLS$_o$ in graph mining is briefly discussed.

The structure of our paper is as follows. In Section 2, we review the background of the problem and related work. In Section 3, we describe our local search algorithm RLS$_o$ for MDS. In Section 4, the multi-start local search algorithm MSRLS$_o$ for MWDS is presented. In Section 5, we present the experimental results and provide a short discussion. Finally, in Section 6, we summarise the contributions and identify open problems.

\section{Background and Related Work}

The problem of finding MDS remains NP-hard also for several very restricted graph classes.
For example, NP-hardness of finding MDS for grids is known, the proof is attributed to Leighton \cite{griddomset}. For unit disk graphs, MDS problem is also NP-hard \cite{unitdisknphard}.
Additionally, Chleb\'{i}k and Chleb\'{i}kov\'{a} have shown that for bipartite graphs with maximum degree $3$ and general graphs with maximum degree $4$, it is NP-hard to approximate MDS within ratios $1 + 1/190$ and $1 + 1/99$, respectively \cite{approxdominatingbounded}.

Regarding the hardness of approximate MDS, approximation ratio \linebreak $\mathcal{O}(\log \Delta)$, where $\Delta$ is the maximum degree of a vertex, can be achieved by the greedy approximation algorithm. However, Feige showed that the logarithmic approximation is the best possible, unless the class NP contains some slightly superpolynomial algorithms \cite{approxsetcover}. Recently, it has also been shown that MDS is hard to approximate within a better than logarithmic ratio for certain graphs with power law degree distribution \cite{powerlawinapprox}.

Exact algorithms require exponential time to solve the problem. To the best of our knowledge, the currently best exact algorithm was proposed by Fomin et al, and finds MDS in $\mathcal{O}(1.5137^n)$ time \cite{complexitydominating}.

\subsection{Greedy approximation algorithm for MDS}

It is known that the greedy approximation algorithm for vertex cover by Chv\'{a}tal \cite{chvatal} can be used to find small dominating sets in polynomial time. This algorithm achieves approximation ratio $H(\Delta)$, where $\Delta$ is the maximum degree of a vertex and $H(n) = \sum_{i=1}^{n} 1/i$ is the $n$-th harmonic number. As we indicated above, $H(\Delta) = \mathcal{O}(\log \Delta)$.
	
The application of the greedy approximation algorithm to find small dominating sets works as follows. For simplicity, we will say that a vertex $v$ is \textit{non-dominated in set $S$} if $v \notin S \wedge \nexists v' \in V [\{v,v'\} \in E \wedge v' \in S]$.	In the algorithm, vertices are ordered based on a value $w(v,S)$, which denotes the \textit{number of non-dominated vertices between the neighbours of the vertex and the vertex itself}. In each iteration, vertex $v$ with the largest $w(v,S)$ in partial dominating set $S$ is taken and put into $S$. The algorithm terminates when $S$ is a dominating set.
\vspace{10pt}

\noindent
The downside of the greedy approximation algorithm is that even though it provides a very good approximation ratio in general, it can overestimate the dominating set size in practice. This holds even for relatively simple graphs such as paths. Therefore, experimental research on heuristics for the classical MDS problem in large graphs should be of a high interest, especially for its applications in large real-world networks.

In contrast to weighted and connected dominating set problems, greedy and distributed \cite{distibuteddominating} approximation algorithms are still among the most popular in MDS. An experimental study of heuristics for MDS was conducted by Sanchis \cite{expdomset}. Performance of algorithms for MDS in real-world networks has been compared by Neh\'{e}z et al \cite{nehez2015comparison}. There has been a recent surge in heuristic algorithms for MWDS \cite{bouamama2016hybrid,wang2017local,lin2016effective}. These algorithms are usually applied to classical benchmarks consisting of graphs inspired by the applications in wireless networks, with up to around $1000$ vertices and with strong results in terms of numerical performance. However, the focus on scalability in the current experimental research on MDS is still somewhat limited.

\subsection{Ant colony optimisation algorithms with local search for MDS}

Hedar and Ismail \cite{gadomset} proposed a hybrid of a genetic algorithm and local search for MDS with a specific fitness function. Another hybrid genetic algorithm was proposed by Potluri and Singh \cite{twoheurdomset}.

An ant colony algorithm for the minimum weight dominating set problem has been proposed by Jovanovic et al \cite{jovanovic2010ant}. Another ant colony optimisation algorithm hybridised with local search (ACO-LS) was proposed by Potluri and Singh \cite{twoheurdomset}. ACO-LS represents the basis for some of the very good heuristics currently available for MDS. Therefore, we will now provide more details on how this algorithm works.

ACO-LS operates on a construction graph, which is a complete graph with vertices corresponding to vertices of our original graph. Vertices of the construction graph are weighted with pheromone, which represents how likely the vertex is to be in the dominating set. ACO-LS first places an initial amount of pheromone $\tau$ uniformly on each vertex. In each iteration, $A$ dominating sets are constructed so that probability of the next vertex being put to the dominating set is proportional to the amount of pheromone on it.

The resulting dominating sets are improved using local search, which iteratively excludes redundant vertices from the dominating set. Redundant vertices are vertices, which can directly be excluded so that the resulting set still remains a dominating set. If there are more redundant vertices to exclude, the vertex is chosen randomly or by choosing the vertex with the lowest degree. The probability of choosing the vertex randomly is $p_r$, the lowest degree vertex is chosen otherwise.

At the end of each iteration, ACO-LS evaporates pheromone on each vertex by a multiplicative factor of $\rho$. Vertices of the smallest dominating set constructed in the current iteration are then reinforced by putting more pheromone on them. The formula that has been used to obtain the updated pheromone value $\tau'$ for vertices in the best dominating set generated in the current iteration is the following:

\begin{equation}
\tau' = \rho \tau + \frac{p_1}{p_2 -f + F}.
\end{equation}

\noindent
This pheromone update rule has originally been used in an ant colony optimisation algorithm for the leaf-constrained minimum spanning tree problem \cite{singh2008new}. In this formula, $f$ is the best dominating set size for the current iteration and $F$ is the best dominating set size found so far. In ACO-LS, the values of parameters $p_1$ and $p_2$ were $p_1 = 1.0$ and $p_2 = 10.0$ \cite{twoheurdomset}. Hence, the vertices used in previously constructed dominating sets will be more likely to occur in the next dominating sets.

In the application of a similar idea to the minimum weight dominating set problem, a preprocessing phase has been added to a similar algorithm called ACO-PP-LS \cite{dominatingweight}. This algorithm extended ACO-LS by adding a routine of generating $M = 100$ maximal independent sets using a greedy algorithm and improving the pheromone values of vertices in these independent sets. These independent sets were constructed by using a list of vertices available for adding to the independent set. In the beginining, a random vertex is added to the independent set. This vertex and its neighbours are then excluded from the list of available vertices. This is process iterated until there are no more vertices available. This preprocessing routine helps the algorithm to start with a probabilistic model that is not entirely random, leading to more rapid convergence for some instances. In ACO-PP-LS, slighly different values of parameters $p_1$ and $p_2$ have also been used. These were $p_1 = 2.0$ and $p_2 = 5.0$. We will use the original parameterisations of ACO-LS \cite{twoheurdomset} and ACO-PP-LS \cite{dominatingweight} in our further experimental investigations.

It is worth mentioning that algorithms based on ant colony optimisation are popular also in other variants of dominating set problems. A similar approach was proposed for the minimum connected dominating set problem \cite{acodominating}.

Because of the inapproximability barriers mentioned above, one can naturally expect that improvements of previous results are dependent on structure of studied graphs. While it can be intractable to provide a significant improvement for an arbitrary graph, for some specific instances this might be possible. In the following, we will tackle the classical MDS by combining a greedy construction procedure mapping a fixed permutation of vertices to a dominating set with an optimisation routine for this permutation.

\section{RLS$_o$: An Order-based Randomised Local Search for MDS}

At this point, we begin with specification of our \textit{order-based randomised local search} (RLS$_o$). The idea to use this approach comes from the field of randomised search heuristics theory \cite{eacomplexity}, in which RLS is studied analytically. Additionally, a similar approach was previously studied in the context of load balancing games \cite{rlsbalancing}.

Our RLS$_o$ algorithm tackles the problem indirectly by searching for a permutation, for which a greedy algorithm constructs as small dominating set as possible. Solution will be represented by a permutation of vertices. This permutation is used as an input to a greedy mapping algorithm, which guarantees that for some permutation, the optimal dominating set will be constructed. This way, the search for minimum dominating set is transformed into search for an optimal permutation.

We now specify the search space, our mapping algorithms and the objective function. Candidate solution $S$ will be now represented as a set $S \subseteq V$, since RLS$_o$ always works with dominating sets. The search space $\Omega$ is represented by the set $S_n$ of all permutations on $n$ vertices, i.e. $\Omega = S_n$.

For the purpose of mapping permutations of vertices to dominating sets, RLS$_o$ will use a simple greedy algorithm, similar to the greedy approximation algorithm. We start with set $S = \emptyset$. The algorithm takes vertices in the ordering determined by permutation $P$. In each iteration $i$, we test whether the current vertex $v_i$ is non-dominated or has any non-dominated neighbours. If $v_i$ or any of its neighbours are non-dominated, then we put $S = S \cup \{v_i\}$, thus, making $v_i$ and all of its neighbours dominated. The algorithm stops when all vertices are dominated, i.e. $S$ is a dominating set.

\begin{table}
\begin{center}
Algorithm 1: Greedy algorithm for mapping of a permutation to a dominating set\vspace{5pt}\\
\begin{tabular}{l|l}
& Input: graph $G = [V,E], V = \{v_1,v_2,...v_n\}$,\\
& permutation $P$ of integers from $\{1,2,...,n\}$\\
& Output: solution $S \subseteq V$\\\hline
1  & $S = \emptyset$\\
2  & for $v \in V$\\
3  & ~~~~$D(v,S) = 1$\\
4  & $i = 0$\\
5  & while $\exists v ~ D(v,S) = 1$\\
6  & ~~~~$v = v_{P_i}$\\
7  & ~~~~if $D(v,S) = 1 \vee \exists v' [\{v,v'\} \in E \wedge D(v',S) = 1]$\\
8  & ~~~~~~~~$S = S \cup \{v\}$\\
9  & ~~~~~~~~$D(v,S) = 0$\\
10 & ~~~~~~~~for $v'$ such that $\{v,v'\} \in E$\\
11 & ~~~~~~~~~~~~$D(v',S) = 0$\\
12 & ~~~~$i = i + 1$\\
13 & return $S$\\
\end{tabular}
\end{center}
\end{table}

In Algorithm 1, we present detailed pseudocode of this greedy mapping algorithm. We define a function $D(v,S)$ such that $D(v,S) = 1$ if $v$ is non-dominated in $S$, i.e. if $w(v,S) \neq 0$, and $D(v,S) = 0$ otherwise. In steps 1-3, we start with an empty set $S$ and set the initial values of $w(v,S)$ for empty $S$. Steps 5-12 represent an iterative procedure. In step 6, we take the $i$-th vertex in the permutation and denote it by $v$. Steps 7-12 are performed only if $v$ is non-dominated or it has a non-dominated neighbour. In step 8, we put $v$ in $S$. In steps 9-11, we set $v$ and all of its neighbours as dominated. The iterative process terminates if all vertices are dominated.

Let $S$ be a dominating set. Then, when mapping dominating sets to permutations of vertices, we will construct permutation $P$ from $S$ by putting the vertices in $S$ first in the permutation. Their order can be arbitrary, we will use ordering by vertex indices. The vertices in $V \backslash S$ are put in $P$ in a uniformly random order after the vertices in $S$.

Consider now how this greedy mapping algorithm behaves when mapping $P$ constructed from $S$ to a resulting dominating set $S'$. The first $|S|$ vertices in $P$ are in $S$, i.e. they form the dominating set. Let $v$ be the current vertex in $P$ to be processed by the greedy algorithm. Vertex $v$ can be non-dominated or have a non-dominated neighbour at the moment when it is processed by iterating over vertices in $P$. In this case, $v$ will be added to $S'$. If $v$ and all of its neighbours are already dominated, it means that some previous vertices were sufficient to dominate $v$ and its neighbours. However, these vertices were already in $S$, i.e. $S$ is not a minimal dominating set and can be improved by excluding $v$. Therefore, at the moment, when first $|S|$ vertices of $P$ are processed, we will have a dominating set $S' \subseteq S$. As a consequence, if $S$ is the minimum dominating set, then a permutation generated by putting the vertices in $S$ first in the permutation must necessarily produce the optimum. Therefore, we can guarantee that there is an optimal permutation for an arbitrary graph.

Since RLS$_o$ operates in the space of dominating sets, the objective will simply be to minimise the cardinality of the dominating set $S$, i.e. the problem is $\min |S|$.

\begin{table}
\begin{center}
Algorithm 2: Order-based randomised local search (RLS$_o$) for MDS\vspace{5pt}\\
\begin{tabular}{l|l}
& Input: graph $G = [V,E]$, lower bound $\gamma_l$ for $\gamma$\\
& Output: solution $S \subseteq V$\\\hline
1  & let $S$ be the result of greedy approximation algorithm on $G$\\
2  & use mapping of dominating sets to permutations to create\\
   & permutation $P$ from $S$\\
3  & while stopping criteria are not met\\
4  & ~~~~$P' = jump(random(2,n),P)$\\ 
5  & ~~~~construct $S'$ from $P'$ using the greedy algorithm\\
   & ~~~~for mapping of a permutation to a dominating set\\
6  & ~~~~if $|S'| \leq |S|$\\
7  & ~~~~~~~~$P = P'$\\
8  & ~~~~~~~~$S = S'$\\
9  & return $S$\\
\end{tabular}
\end{center}
\end{table}

RLS$_o$ will now search for a permutation of vertices, for which the greedy mapping from Algorithm 1 produces as small dominating set as possible. As an elementary move, RLS$_o$ will use the $jump$ perturbation operator, which works as follows.

Let $j$ be an integer from $\{2,3,...,n\}$. Then, perturbation operator \linebreak $jump(j,P)$, will take the element at position $j$ in permutation $P$ and put it into position $1$, i.e. to the front. The elements formerly between positions $1$ and $j-1$ will then be shifted one position to the right. The returned result is this new permutation.

In Algorithm 2, we present the pseudocode of RLS$_o$. In step 1, we start with a dominating set $S$ constructed by the greedy approximation algorithm. In step 2, we use the mapping of dominating sets to permutations of vertices to create an initial permutation $P$. Next, we perform an iterative procedure. In each iteration, we verify whether $S$ is optimal by checking its size against a lower bound $\gamma_l$ that can be provided as an input, if it is known for the instance. This lower bound is used as a stopping criterion, since the search can be stopped if a known optimum has been found. The other stopping criterion used in our experiments below was a fixed time limit. In step 4, we perform $jump(random(2,n),P)$, i.e. we take a uniformly random vertex (except the currently first one) and put it to the first position in the permutation to construct $P'$. In step 5, we use the greedy mapping from Algorithm 1 to construct new dominating set $S'$ from $P'$. In steps 6-8, $P'$ and $S'$ are accepted as the new $P$ and $S$, if $S'$ is at least as good as $S$.

\section{MSRLS$_o$: A Multi-start Order-based Randomised Local Search for MWDS}

RLS$_o$ is an algorithm that is well-suited for MDS in large social and complex networks, since it preserves the logarithmic approximation and provides a well-scalable routine to optimise the dominating sets. However, a large body of current literature is dedicated to the minimum weight dominating set problem (MWDS) in which the aim is to find the dominating set with minimum total weight in a graph with weighted vertices \cite{wang2017local,lin2016effective,chaurasia2015hybrid}.

MWDS is usually solved for benchmark graphs derived from wireless network applications \cite{jovanovic2010ant}. This leads to moderately large instances that have multiple attractors in the search space. This is why a multi-start variant of RLS$_o$ is required to obtain good results for MWDS. In the following, we propose such an extension of RLS$_o$ and we denote it by MSRLS$_o$.

\begin{table}
\begin{center}
Algorithm 3: Multi-start Order-based randomised local search (MSRLS$_o$) for MWDS\vspace{5pt}\\
\begin{tabular}{l|l}
& Input: graph $G = [V,E]$\\
& maximum number of iterations without improvement $i_{max}$\\
& extended maximum number of iterations without improvement $i_{best}$\\
& Output: solution $S \subseteq V$\\\hline
1  & with probability $p_{gr}$\\
2  & ~~~~let $S$ be the result of greedy algorithm for minimum\\
	 & ~~~~weight dominating set on $G$\\
3  & ~~~~use mapping of dominating sets to permutations\\
   & ~~~~to create permutation $P$ from $S$\\
4  & otherwise\\
5  & ~~~~let $P$ be a random permutation of vertices in $V$\\
6  & construct $S$ from $P$ using the greedy mapping algorithm\\
7  & $S_{best} = S$, $i = 0$, $ext = false$\\
8  & while stopping criteria are not met\\
9  & ~~~~if $ext = false \wedge i > i_{max} \vee ext = true \wedge i > i_{best}$\\
10  & ~~~~~~~~restart the search constructing $P$ and $S$ using steps 1-5\\
11  & ~~~~~~~~$i = 0$, $ext = false$\\
12  & ~~~~$P' = jump(random(2,n),P)$\\ 
13  & ~~~~construct $S'$ from $P'$ using the greedy mapping algorithm\\
14  & ~~~~if $weight(S') \geq weight(S)$ $i = i+1$ else $i = 0$\\
15  & ~~~~if $weight(S') \leq weight(S)$\\
16  & ~~~~~~~~$P = P'$\\
17  & ~~~~~~~~$S = S'$\\
18  & ~~~~~~~~if $weight(S_{best}) > weight(S)$ $S_{best} = S$, $ext = true$\\
19  & return $S_{best}$\\
\end{tabular}
\end{center}
\end{table}

The pseudocode of MSRLS$_o$ is outlined in Algorithm 3. Compared to RLS$_o$, a slightly different greedy algorithm is used in the initialisation procedure of MSRLS$_o$. With probability $p_{gr}$, we construct the initial solution $S$ using a greedy approach, choosing the vertex with highest priority $d_S(v) / w(v)$, where $d_S(v)$ is the number of currently non-dominated neighbours of vertex $v$ and $w(v)$ is its weight \cite{bouamama2016hybrid,chvatal}. The vertices in $S$ are then put at the start of initial permutation $P$ and the other vertices are ordered uniformly at random, similarly to RLS$_o$ for MDS. Otherwise, with probability $1-p_{gr}$, the algorithm starts with a random permutation $P$. This process is summarised in steps 1-5 of Algorithm 3.

MSRLS$_o$ performs repeated sampling routines of promising weighted dominating sets. Let one such routine be called a cycle. If $i_{max}$ iterations without improvement are reached, the algorithm simply restarts the search. However, if an improvement of the best weighted dominating set found so far is achieved, then the current cycle is extended to $i_{best}$ iterations. This is reflected in steps 9-11 of Algorithm 3. A maximum of $c_{max}$ cycles are carried out in MSRLS$_o$. The $jump$ operator is used in step 12. The decision of whether the new permutation and solution are accepted is carried out in steps 15-18. Note that the total weight of dominating set $S$ is $weight(S) = \sum_{v \in S} weight(v)$.

\section{Experimental Evaluation}

In this section, we present the experimental results of RLS$_o$ and its comparison to ACO-LS, ACO-PP-LS, as well as the greedy approximation algorithm.  We also present the experimental results of MSRLS$_o$ for MWDS. In addition, we experiment with our own modification of ACO-LS, an algorithm we denote by ACO-LS-S, which works much better than the original ACO-LS in solving MDS for large sparse graphs.

The idea behind ACO-LS-S is that for large graphs, it is preferable to avoid random walks on complete construction graphs to increase the scalability of the algorithm, even if it means a decrease in numerical performance as a tradeoff.

We first present an overview of the experimental settings and structure of the evaluation. In accordance with previous studies on both ACO-LS and hybrid genetic algorithms \cite{dominatingweight,twoheurdomset,gadomset}, we used unit disk graphs as test instances.  Next, we present results for random scale-free networks generated by Barab\'{a}si-Albert (BA) model \cite{barabasialbert2}. Finally, results are presented for a wide spectrum of real-world graphs, including samples from two social network services, graphs studied in network science, as well as several DIMACS graphs. Particular emphasis is put on scalability to large problem instances. An illustration of the relation of our work to graph mining and small-world network properties is also given.

\subsection{Experimental Settings}

We conducted a series of experiments for both unit disk graphs, which were used in previous studies, as well as artificial and real-world complex networks. This will allow a comparison both in settings similar to previous studies \cite{dominatingweight,twoheurdomset,gadomset} and provide a new perspective on experimental algorithms for MDS. The graphs were represented using adjacency lists, as many of the studied graphs are relatively large but sparse.

We first performed experiments for randomly generated unit disk graphs, which represent a model of wireless networks. A unit disk graph is an intersection graph, in which vertices correspond to points in a certain area. The points represent omnidirectional antennas and an edge represents that the ranges of two antennas overlap. In our experiments, we will use a square area of size $M \times M$ and antennas are placed uniformly at random. An edge is put between two vertices if their distance is at most $range$. This is equivalent to a setting used by Potluri and Singh \cite{twoheurdomset}.

Next, we present experimental results for BA model, which represents a canonical model of growing scale-free networks \cite{barabasialbert2}. BA model is particularly interesting for our investigations, since it generates networks with power law degree distribution, which is typical for many real-world networks and allows that both number of vertices and their ``connectivity'' are tuned.

In the largest part of our experiments, we test the algorithms in solving MDS
for various types of real-world graphs\footnote{The information about the network data and the corresponding links are available at:\\
\url{http://davidchalupa.github.io/research/data/networks.html}.}. These include data obtained from two different social networks - graphs obtained from the publicly available circles data from \textit{Google+}, and samples of social network \textit{Pokec}, which has been previously analyzed in large scale \cite{pokecanalysis}. These networks have up to $50000$ vertices, i.e. scalability will be a crucial issue. We enhance the benchmark by several networks studied in network science \cite{communitystructure1,lesmis,adjnoun,zachary} and coappearance networks of classical literary works' characters used in DIMACS graph colouring benchmark \cite{johnson::dimacs}.

We reimplemented both the greedy approximation algorithm and ACO-LS. The greedy approximation algorithm will be denoted by GREEDY. Apart from the original variants of ACO-LS and ACO-PP-LS, we experimented with our ACO-LS-S modification of the ant-based framework, which allows transitions only between adjacent vertices in our graph. We will shortly see that this leads to an increase in scalability to large graphs.

ACO-LS and ACO-PP-LS perform random walks on complete weighted construction graphs. With ACO-LS-S, our experiments focused on constraining the transitions only to vertices, which are adjacent in the original graph. Apart from this, there is only one further difference between ACO-LS and ACO-LS-S. ACO-LS-S first performs a run of GREEDY and the amount of pheromone value $1000.0$ is placed on the vertices in the dominating set constructed by GREEDY. This is a somewhat similar idea to what ACO-PP-LS uses as a preprocessing routine. Other vertices have initial pheromone value $10.0$ used in ACO-LS.

GREEDY was repeated $1000$ times for each instance with a randomised setting, i.e. ties are broken uniformly at random. For ACO-LS, we used the same values of parameters as in the original paper \cite{twoheurdomset}. We used $A = 20$ constructed dominating sets per iteration, pheromone evaporation rate $\rho = 0.985$, probability $p_r = 0.6$ of excluding redundant vertices randomly and the initial pheromone value was $\tau_0 = 10.0$. Pheromone update parameters were $p_1 = 1.0$ and $p_2 = 10.0$ for ACO-LS. ACO-PP-LS used the same parameter configuration, apart from the pheromone update parameters, which were $p_1 = 2.0$ and $p_2 = 5.0$. Parameters of ACO-LS-S were equivalent to those of ACO-LS, except the initial reinforcement of pheromone for vertices used by GREEDY as indicated above. In the cases of ACO-LS, ACO-PP-LS, ACO-LS-S and RLS$_o$, we stopped whenever a lower bound on the size of MDS was reached (the methodology of finding the lower bound is described below) or the algorithm has reached the time limit. Time limits were specific for each series of experiments. All experiments were performed on a machine with Intel Core i7-5960X 3 GHz CPU with 64 GB RAM and a code written in C++, compiled with 32-bit MinGW compiler under Windows 8.1 with -O3 optimisation option. 

In the experiments for real-world graphs, we also include a lower bound for the domination number $\gamma$, which has been used as a stopping criterion to verify if optimum has been found. This value is computed as a solution to the linear programming relaxation of MDS. Let $S \subseteq V$ and let $x_i \in \{0,1\}$ be a variable such that vertex $x_i = 1$ if $v_i \in S$ and $x_i = 0$ otherwise. Then, MDS can be formulated as a minimisation of value $\sum_{i=1}^{n} x_i$ subject to $x_i + \sum_{\{v_i,v_j\} \in E} x_j \geq 1$ for all $i = 1,2,...,n$. By relaxing the constraint that $x_i$ is binary and assuming that $0 \leq x_i \leq 1$, one obtains a simple linear programming problem that can be solved in polynomial time to provide a very good lower bound. We have used the COIN-OR package \cite{BonamiAlgorithmicMixedIntegerPrograms,LinderothMilp} to find these lower bounds for each real-world graph.

\subsection{Results of RLS$_o$ in MDS for Unit Disk Graphs}

Unit disk graphs were created by putting $n$ points randomly in a square area of size
$M \times M$. Values of $n$ equal to 50, 100, 250, 500, 750 and 1000 were used. For 50, 100 and 250 vertices, we set $M = 1000$ and for 500, 750 and 1000, we set $M = 2000$. For each value of $n$, three different values of $range$ were used.
We obtained 10 unit disk graphs using these parameters, for which we computed minimum, maximum and average values for 1000 runs of GREEDY for each graph, and average values of dominating set sizes for ACO-LS, ACO-PP-LS, ACO-LS-S and RLS$_o$. Each run of ACO-LS, ACO-PP-LS, ACO-LS-S and RLS$_o$ was stopped after 3 minutes. The time limit was chosen as a more fair criterion than maximum number of iterations, since the computational cost of one iteration for RLS$_o$ is lower than for ACO-LS and ACO-PP-LS. Additionally, the numbers of dominating sets evaluated by ACO-LS in our experiments consistently exceed $2 \times 10^4$. This was the limit used to obtain the original results of ACO-LS \cite{twoheurdomset}, which assures that the comparison presented in this paper is reasonably fair.

\begin{table}
\caption{Experimental results of the studied algorithms for unit disk graphs, presenting the average size of dominating set found, as well as the average number of evaluations in thousands performed within the time limit of $3$ minutes per run.}
{\tiny
\begin{center}
\begin{tabular}{l l l l l l l l l l l l}\toprule
$[n,range]$	& \multicolumn{3}{l}{Greedy}			& \multicolumn{2}{l}{ACO-LS}		& \multicolumn{2}{l}{ACO-PP-LS}	& \multicolumn{2}{l}{ACO-LS-S}	& \multicolumn{2}{l}{RLS$_o$} \\
			& min	& max	& $E[|S|]$	& $E[|S|]$		& eval			& $E[|S|]$		& eval			& $E[|S|]$ 	& eval & $E[|S|]$		& eval \\
			& 			& 			&		&			& $\times 10^3$	&			& $\times 10^3$	& 			& $\times 10^3$ & 			& $\times 10^3$ \\\midrule
$[50,150]$		& 15.2			& 16.1	& 15.6	& \textbf{14.6}	& 14748		& \textbf{14.6}	& 16366		& \textbf{14.6}		& 17212		&	\textbf{14.6}		& 226389 \\
$[50,200]$		& 10.6			& 12		& 11.1	& \textbf{10.1}	&	19455		& \textbf{10.1}	& 22736		& \textbf{10.1}		& 20339		&	\textbf{10.1}		& 195807 \\
$[50,250]$		& 7.5				& 8.1		& 7.7		& \textbf{7.1}		& 24652		& \textbf{7.1}		& 30412		& 7.2			& 20174		&	\textbf{7.1}		& 190241 \\
$[100,150]$		& 18.3			& 20.8	& 19.5	& 17.5	& 5782		& 17.5	& 6614		& 17.5		& 6333		&	\textbf{17.4}		& 122087 \\
$[100,200]$		& 11.9			& 13.9	& 12.8	& \textbf{10.7}	& 8357		& \textbf{10.7}	& 10314		& 11.3		& 7835		&	\textbf{10.7}		& 113578 \\
$[100,250]$		& 8.5				& 10.1	& 9.2		& \textbf{7.5}		& 10473		& \textbf{7.5}		& 13216		& 7.9			& 9526		&	\textbf{7.5}		& 98837 \\
$[250,150]$		& 21.6			& 25.2	& 23.2	& 18.3	& 1933		& 18.2	& 2543		& 20.2		& 1772		&	\textbf{18}			& 47323 \\
$[250,200]$		& 13.5			& 15.7	& 14.5	& 11.1	& 2515		& 11.2	& 3480		& 12.2		& 2798		&	\textbf{11}			& 33063 \\
$[250,250]$		& 10.3			& 12.3	& 11.2	& \textbf{8}			& 2755		& \textbf{8}			& 3763		& \textbf{8}				& 2906		&	\textbf{8}			& 25613 \\
$[500,150]$		& 71.9			& 81		& 76.2	& 64.5	& 354			& 64.5	& 440			& 69.9		& 316			&	\textbf{63.8}		& 34168 \\
$[500,200]$		& 45.6			& 53.4	& 49.1	& 39.8	& 493			& 39.8	& 673			& 45.4		& 400			&	\textbf{38.6}		& 24822 \\
$[500,250]$		& 31.4			& 36.9	& 33.7	& 26.4	& 627			& 26.5	& 894			& 30.7		& 753			&	\textbf{25.8}		& 17502 \\
$[750,150]$		& 77.8			& 88.1	& 82.5	& 67.3	& 191			& 67.4	& 256			& 78.6		& 176			&	\textbf{65.2}		& 17329 \\
$[750,200]$		& 47.9			& 55.3	& 51.4	& 40.4	& 260			& 40.5	& 390			& 48.3		& 342			&	\textbf{38.9}		& 11053 \\
$[750,250]$		& 32.2			& 38.7	& 35.3	& 26.8	& 345			& 27.2	& 537			& 32.3		& 470			&	\textbf{26.3}		& 9579 \\
$[1000,150]$	& 80.2			& 90.2	& 85		& 69.3	& 116			& 69.3	& 169			& 82.1		& 89			&	\textbf{66.7}		& 9141 \\
$[1000,200]$	& 49.5			& 57.9	& 53.6	& 42.2	& 161 		& 42		& 249			& 50.2		& 260			&	\textbf{40.4}		& 7453 \\
$[1000,250]$	& 34.3			& 40.4	& 37		& 27.6	& 215			& 27.6	& 344			& 32.9		& 292			&	\textbf{27}			& 6101 \\
\bottomrule
\end{tabular}
\end{center}
}
\end{table}

In Table 1, the results of the algorithms are presented. The first column contains the unit disk graph parameters. The next three columns contain results of GREEDY, including minimum, maximum and average dominating set sizes. Results of ACO-LS, ACO-PP-LS, ACO-LS-S and RLS$_o$ are presented in the next columns, including the average obtained value and the numbers of evaluated dominating sets (in the cases of ACO-LS and ACO-LS-S, this is the number of dominating sets generated after the LS phase). Best results are highlighted in bold.

One can see that the results of GREEDY lag behind the results of ACO-LS, ACO-PP-LS, ACO-LS-S and RLS$_o$, even when restricted to the best runs. ACO-LS and ACO-PP-LS provided very similar results, even though it is worth pointing out that working with a smaller population of ants made ACO-PP-LS perform better than ACO-LS. For smaller unit disk graphs, ACO-LS, ACO-PP-LS and RLS$_o$ perform comparably well. However, the difference in performance between the ant-based algorithms and RLS$_o$ starts to be visible for larger graphs with $n \geq 500$, for which RLS$_o$ consistently provided better results than alternative approaches. The number of evaluations performed by RLS$_o$ within the time limit was also much higher than those by ACO-LS, ACO-PP-LS and ACO-LS-S. RLS$_o$ constructs the dominating set in a way such that no elimination of redundant vertices is needed, leading to more compact intermediate dominating sets and a more rapid search strategy. However, it is also worth pointing out that RLS$_o$ could potentially be used as a local search subroutine within the framework of ACO. This would likely lead to an algorithm with a very strong tradeoff between performance and scalability.

\subsection{Results of RLS$_o$ in MDS for Scale-free Networks}

In addition to unit disk graphs, we used BA model to generate artificial scale-free networks for further experiments. In each time step in BA model, one vertex comes to the network and brings $w$ edges, which are attached preferentially, i.e. probability of attachment to a vertex is proportional to its current degree. For example, in a social network, this rule is interpreted in a way that a person with a higher number of contacts is likely to get even more contacts. Therefore, we used BA model to evaluate how well studied algorithms perform for networks, which follow this rule. The initial graph was simply a path on $w$ vertices. We stopped whenever the resulting graph had $n$ vertices. Hence, a particular instance is defined by parameters $n$ and $w$.

Table 2 presents results obtained for the random scale-free networks. The structure of the table is almost identical to the structure used for unit disk graphs, with first column presenting the pair $[n,w]$, which determines model parameters. Similarly, the results are averaged over 10 instances for the specified parameter values.

Interestingly, ACO-LS, ACO-PP-LS, ACO-LS-S and RLS$_o$ exhibit similar specificities, as observed in experiments for unit disk graphs. For smaller instances, ACO-PP-LS and RLS$_o$ perform comparably. Based on these results, RLS$_o$ appears to provide better results for instances generated with higher values of $w$, which are denser graphs with a slightly higher number of triangles \cite{bollobas}.

Additionally, for large instances with $2000$ vertices, we observed that ACO-PP-LS performs better than ACO-LS, which is not the case for smaller graphs. ACO-LS-S is able to evaluate more candidate solutions than ACO-LS and ACO-PP-LS. However, this does not translate into a good numerical performance. This unfortunately highlights the fact that the quadratically complex construction routine of ant-based algorithms is crucial for numerical performance on some types of instances. On the other hand, RLS$_o$ needs $\mathcal{O}(m)$ time to construct a dominating set, where $m$ is the number of edges. This has contributed to the success of RLS$_o$, potentially highlighting its strong role in future designs of hybrid algorithms for MDS.

\begin{table}
\caption{Experimental results of the studied algorithms for graphs generated by BA model, presenting the average size of dominating set found, as well as the average number of evaluations in thousands performed within the time limit of $3$ minutes per run.}
{\tiny
\begin{center}
\begin{tabular}{l l l l l l l l l l l l}\toprule
$[n,w]$		& \multicolumn{3}{l}{GREEDY}			& \multicolumn{2}{l}{ACO-LS}		& \multicolumn{2}{l}{ACO-PP-LS}		& \multicolumn{2}{l}{ACO-LS-S}	& \multicolumn{2}{l}{RLS$_o$} \\
			& min	& max	& $E[|S|]$	& $E[|S|]$		& eval			& $E[|S|]$		& eval			& $E[|S|]$ 	& eval & $E[|S|]$		& eval \\
			& 			& 					&		&			& $\times 10^3$	&			& $\times 10^3$	& 			& $\times 10^3$ & 			& $\times 10^3$ \\\midrule
$[500,2]$			& 95.2			& 101.7	& 97.9	& \textbf{93.7}	& 308			& \textbf{93.7}	& 363			& 94.2		& 485			&	\textbf{93.7}		& 35917 \\
$[500,3]$			& 68.2			& 75.2	& 71.1	& 66.3	& 409			& 66.6	& 528			& 69.6		& 576			&	\textbf{66.2}		& 34412 \\
$[500,4]$			& 55.1			& 61.4	& 57.9	& 53.4	& 477			& 53.6	& 636			& 58			& 597			&	\textbf{53.2}		& 32707 \\
$[1000,2]$		& 186.3			& 197.6	& 191.5	& \textbf{182.4}	& 70			& 182.5	& 87			& 187.6		& 118			&	\textbf{182.4}		& 15229 \\
$[1000,3]$		& 138.9			& 149.4	& 144		& 133.5	& 87			& 134.1	& 116			& 148.5		& 124			&	\textbf{132.9}		& 13904 \\
$[1000,4]$		& 108.9			& 119.5	& 113.6	& 105.5	& 101			& 106.2	& 146			& 121.4		& 120			&	\textbf{105}			& 12900 \\
$[2000,2]$		& 373.9			& 388.1	& 381.1	& 385.9	& 6				& 370.4	& 9				& 384.3		& 27			&	\textbf{362.5}		& 4655 \\
$[2000,3]$		& 276.4			& 291.6	& 283.4	& 301.6	& 6				& 282		& 10			& 301.9		& 25			&	\textbf{263.5}		& 4428 \\
$[2000,4]$		& 215.6			& 230.8	& 222.7	& 241.7	& 7				& 221.7	& 11			& 249.3		& 24			&	\textbf{205.1}		& 4327 \\
\bottomrule
\end{tabular}
\end{center}
}
\end{table}

\subsection{Results of RLS$_o$ in MDS for Real-world Graphs}

In this section, we present the results of our algorithm for several real-world graphs. We will explore the performance of RLS$_o$, ACO-LS, ACO-PP-LS, ACO-LS-S and GREEDY for \textit{real-world data from social networks Google+ and Pokec}, several \textit{graphs studied in network science} and coappearance networks from \textit{DIMACS graphs}.

Similarly to previous experiments, each run of GREEDY was repeated $1000$ times and each run of ACO-LS, ACO-PP-LS, ACO-LS-S, RLS$_o$ and MSRLS$_o$ was repeated $20$ times. For ACO-LS, ACO-PP-LS, ACO-LS-S and RLS$_o$, we were searching for the smallest dominating set sizes obtained, i.e. the upper bounds for domination numbers computed by each of these algorithms. Time limits for these experiments were higher, since we have tested the algorithms for several larger graphs with up to $50000$ vertices. Each run of ACO-LS, ACO-PP-LS, ACO-LS-S and RLS$_o$ was stopped after $10$ minutes or when the lower bound for the minimum dominating set size has been reached, computed from the linear programming relaxation of MDS, i.e. an optimum has been found. For each instance, we also present the results of long-running experiments with $60$ minute time limit, to determine how good near-optimal solutions the algorithms can compute. These experiments were repeated $10$ times.

\begin{table}
\caption{Experimental results of the studied algorithms for samples of social networks \textit{Google+} and \textit{Pokec} sampled from $20$ runs with time limit of $10$ minutes.}
{\tiny
\begin{center}
\begin{tabular}{l l l l l l l l l l l}\toprule
graph	& \multicolumn{2}{l}{GREEDY}	& \multicolumn{2}{l}{ACO-LS}	& \multicolumn{2}{l}{ACO-PP-LS}			& \multicolumn{2}{l}{ACO-LS-S}			& \multicolumn{2}{l}{RLS$_o$} \\
							& min			& avg			& min		& avg		& min		& avg		& min		& avg		& min		& avg	 \\\midrule
\multicolumn{6}{l}{Samples from \textit{ Google+}*}\\\midrule
$gplus\_500$, $\gamma =  42$		& \textbf{42}	&	42.08			&	\textbf{42}	&	42	&	\textbf{42}	&	42	& \textbf{42}	&	42	& \textbf{42} &	42	\\			
$gplus\_2000$, $\gamma =  170$		& 175		&	177.38				& 	\textbf{170}	&	170		&	\textbf{170}	&	170	& \textbf{170}	&	170	& \textbf{170} &	170 \\				
$gplus\_10000$, $\gamma =  861$	& 890		&	896					& 1083		&	1112.8	&	1088 &	1110.9	& \textbf{861}	 &	862		& \textbf{861} &	861.2	 \\															
$gplus\_20000$, $\gamma  \geq 1716$	& 1799		&	1810.8	& 2202			&	2264.1	&	2206	&	2253	& 1725 		&	1726	& \textbf{1720} &	1724.4	\\									
$gplus\_50000$, $\gamma  \geq 4566$	& 4819		&	4840.19	& 5908			&	5988.7	&	5909	&	6002.9	& \textbf{4631}		&	4633.9	& 4653 &	4671.4	\\\midrule					
\multicolumn{6}{l}{Samples from \textit{ Pokec}*}\\\midrule
$pokec\_500$, $\gamma =  16$		& \textbf{16}	&	16				&	\textbf{16}	&	16	 &	\textbf{16}	&	16	  & \textbf{16}	&	16	 & \textbf{16} &	16  \\		
$pokec\_2000$, $\gamma =  75$		& \textbf{75}	&	75				& 	\textbf{75}	&	75		 &	\textbf{75}	&	75	 & \textbf{75}	&	75	 & \textbf{75} &	75 	  \\	
$pokec\_10000$, $\gamma =  413$	& \textbf{413}	&	413.1	& 590	 		&	623.3	 & 591	&	620.3	 & \textbf{413}	&	413	 & \textbf{413} &	413 \\								
$pokec\_20000$, $\gamma =  921$	& 922		&	926.2					& 1382			&	1442.3	 &	1348	&	1439.1		& \textbf{921} &	921	& \textbf{921}	&	921.6	 \\										
$pokec\_50000$, $\gamma  \geq 2706$	& 2757	&	2775.9		& 4095	&	4171.1	& 4071	&	4178.8	& \textbf{2727} 	&	2733		& 2735	&	2743.7 \\\bottomrule				
\end{tabular}
\end{center}
{\scriptsize * All of these network samples are publicly available at:\\
\url{http://davidchalupa.github.io/research/data/social.html}.}

}
\end{table}

\begin{table}
\caption{The best experimental results of the studied algorithms for samples of social networks \textit{Google+} and \textit{Pokec} sampled from $10$ long runs with time limit of $1$ hour.}
{\tiny
\begin{center}
\begin{tabular}{l l l l l}\toprule
graph							& ACO-LS		& ACO-PP-LS		& ACO-LS-S	& RLS$_o$ \\
								&		&			& 			& \\\midrule
\multicolumn{5}{l}{Samples from \textit{ Google+}*}\\\midrule
$gplus\_500$, $\gamma = 42$				& \textbf{42}	&		\textbf{42}	& \textbf{42}	& \textbf{42} \\			
$gplus\_2000$, $\gamma = 170$		& \textbf{170}	&	\textbf{170}	& \textbf{170}	& \textbf{170} \\				
$gplus\_10000$, $\gamma = 861$	& 1061		&		979 & 862	 	& \textbf{861} \\															
$gplus\_20000$, $\gamma \geq 1716$	& 2214		&	2206	& 1725 		& \textbf{1717} \\									
$gplus\_50000$, $\gamma \geq 4566$	& 5926		&	5923	& 4632		& \textbf{4585} \\					
\midrule
\multicolumn{5}{l}{Samples from \textit{ Pokec}*}\\\midrule
$pokec\_500$, $\gamma = 16$		& \textbf{16}	&	\textbf{16} & \textbf{16}	& \textbf{16} \\		
$pokec\_2000$, $\gamma = 75$		& \textbf{75}	&	\textbf{75}	& \textbf{75}	& \textbf{75} \\	
$pokec\_10000$, $\gamma = 413$	& 563 		& 543	& \textbf{413}	& \textbf{413} \\								
$pokec\_20000$, $\gamma = 921$	& 1401		 &	1381	& \textbf{921} & \textbf{921} \\										
$pokec\_50000$, $\gamma \geq 2706$	& 4085		& 4104	& 2726 		& \textbf{2714} \\\bottomrule				
\end{tabular}
\end{center}
}
\end{table}

Table 3 presents results obtained for samples from social networks \textit{Google+} and \textit{Pokec} and Table 4 presents the results obtained in the long runs. These social network samples have $500$, $2000$, $10000$, $20000$ and $50000$ vertices, respectively. Table contains the minimum and maximum value of dominating set size obtained by GREEDY and minimum and average sizes of dominating sets obtained by ACO-LS, ACO-PP-LS, ACO-LS-S and RLS$_o$. The best results obtained are reported for the long runs in Table 4. For smaller graphs, ACO-LS, ACO-PP-LS, ACO-LS-S and RLS$_o$ perform comparably. For graphs with $10000$ vertices, ACO-LS-S and RLS$_o$ start to stand out due to their scalability advantage. ACO-LS-S restricts the random walks only to the original sparse graph. However, ACO-LS-S still performed the LS subroutine eliminating the redundant vertices, which suggests that the scalability of ant-based algorithms is influenced mainly by the random walk on the construction graph, rather than the redundant vertex elimination. It is worth mentioning that this is partly affected by implementation techniques used. One can also notice that ACO-LS-S and RLS$_o$ performed competitively in the shorter runs. However, RLS$_o$ outperformed ACO-LS-S when given more computational time.

\begin{table}
\caption{Experimental results of the studied algorithms for network science instances and DIMACS instances sampled from $20$ runs with time limit of $10$ minutes.}
{\tiny
\begin{center}
\begin{tabular}{l l l l l l l l l l l}\toprule
graph			& \multicolumn{2}{l}{GREEDY}	& \multicolumn{2}{l}{ACO-LS}	& \multicolumn{2}{l}{ACO-PP-LS}			& \multicolumn{2}{l}{ACO-LS-S}			& \multicolumn{2}{l}{RLS$_o$} \\
							& min			& avg			& min		& avg		& min		& avg		& min		& avg		& min		& avg	  \\\midrule
\multicolumn{6}{l}{Graphs from Newman's network data repository*}\\\midrule
\textit{adjnoun} \cite{adjnoun}, $\gamma =  18$				& \textbf{18}	& 	18.8		& \textbf{18}			&	18		 &		\textbf{18}	&	18	& \textbf{18}	&	18	& \textbf{18} &	18		 \\	
\textit{football} \cite{communitystructure1}, $\gamma =  12$		&	13		& 14.3			& \textbf{12}		&	12.1		&		\textbf{12}	&	12.8	& 13			&	13	& \textbf{12} &	12		\\			
\textit{lesmis} \cite{lesmis}, $\gamma =  10$					& \textbf{10}	& 10.5		& \textbf{10}	&		10	&		\textbf{10}	&	10	& \textbf{10}	&	10	& \textbf{10} &	10		 \\				
\textit{netscience} \cite{adjnoun}, $\gamma =  477$			& \textbf{477}	& 478		& 	\textbf{477}	&	477	&		\textbf{477}	&	477	& \textbf{477}	&	477.4	& \textbf{477} &	477		\\					
\textit{zachary} \cite{zachary}, $\gamma =  4$				& \textbf{4}	&  4	& \textbf{4}	&  4		&	\textbf{4} &	4	& \textbf{4}	&	4	& \textbf{4}	&		4	\\					
\textit{celegansneural} \cite{wattsstrogatz}, $\gamma =  16$		& 17		& 18				& \textbf{16}	&	16		&	\textbf{16}	&	16		& \textbf{16}	&	16.5	& \textbf{16}	& 16	 \\	
\textit{dolphins} \cite{dolphins}, $\gamma =  14$				& 15			& 15.6			& \textbf{14}		&	14	&	\textbf{14}		&	14	& \textbf{14}	&	14	& \textbf{14} &		14	\\					
\textit{astro-ph} \cite{collaborations}, $\gamma \geq 2928$		& 3011		& 3024.2		& 3145		&	3157.8	&		3139	&	3053.9	& 3045		&	3053.9	& \textbf{2930} &	2930.4	 \\							
\textit{cond-mat} \cite{collaborations}, $\gamma \geq 3393$		& 3445		& 3454.6		& 3707	&	3727.3	&		3686	&	3722.9	& 3501 		&	3506.4	& \textbf{3394} &	3394	 \\							
\textit{cond-mat-2003} \cite{collaborations},  $\gamma \geq 5377$	& 5489		& 5507.2		& 5974	&	5994		&	5955	&	5999		& 5610	&	5625.7	& \textbf{5382} &	5385	\\					
\textit{cond-mat-2005} \cite{collaborations},  $\gamma \geq 6507$	& 6636		& 6652.1		& 7284	&	7317.5		&	7280	&	7309.4	& 6831	&	6847.1		& \textbf{6519} &	6527.2	\\						
\textit{hep-th} \cite{collaborations},  $\gamma \geq 2612$			& 2633		&	2642	& 2767		&	2776.4	&		2768	&	2776.4	& 2650 	&	2658.3		& \textbf{2613}	&	2613	 \\								
\textit{power} \cite{wattsstrogatz},  $\gamma \geq 1472$			& 1545		& 1579		& 1623		 		&	1632.7	&		1588	&	1602.4	& 1567		&	1578.7	& \textbf{1481}	&	1481 \\								
\textit{as-22july06}, $\gamma =  2026$**						& 2028		& 2030		& 2190			&	2239.5	&	2216		&	2247.2	& \textbf{2026} &	2026.7	& \textbf{2026} &	2026	 \\											
\textit{polbooks}, $\gamma =  13$**							& 14			& 14.5			& \textbf{14}		&	14	&	\textbf{13}		&	13		& \textbf{13}	&	13	& \textbf{13} &	13		\\								
\midrule
\multicolumn{6}{l}{DIMACS graphs*** \cite{johnson::dimacs}}\\\midrule
\textit{anna}, $\gamma =  12$								& \textbf{12}	& 12	& \textbf{12}	&	12	& \textbf{12}		&	12	& \textbf{12}	&	12	& \textbf{12} &	12		\\							
\textit{homer}, $\gamma =  96$							& \textbf{96}	& 96.3	& \textbf{96}	&	96		&		 \textbf{96}	&	96	& \textbf{96}	&	96	& \textbf{96} &	96		\\								
\textit{david}, $\gamma =  2$								& \textbf{2}	& 2	& \textbf{2}	&	2	&		\textbf{2}	&	2	& \textbf{2}	&	2	& \textbf{2} &	2		\\								
\textit{huck}, $\gamma =  9$								& \textbf{9}	& 9	& \textbf{9}	&	9	&	\textbf{9}		&	9	& \textbf{9}	&	9	& \textbf{9}	&		9		\\\bottomrule		
\end{tabular}
\end{center}
{\scriptsize * They are published in Newman's network data repository:\\
\url{http://www-personal.umich.edu/~mejn/netdata/}.\\
*** Snapshot of the Internet and the political books network have not been published in a research paper.\\
**** DIMACS graphs are also available online:\\
\url{http://mat.gsia.cmu.edu/COLOR/instances.html}
}
}
\end{table}

\begin{table}
\caption{The best experimental results of the studied algorithms for network science instances and DIMACS instances sampled from $10$ long runs with time limit of $1$ hour.}
{\tiny
\begin{center}
\begin{tabular}{l l l l l}\toprule
graph											& ACO-LS		&	ACO-PP-LS	& ACO-LS-S	& RLS$_o$ \\
												&			&			&			& \\\midrule
\multicolumn{5}{l}{Graphs from Newman's network data repository}\\\midrule
$adjnoun$ \cite{adjnoun}, $\gamma = 18$					& \textbf{18}	&		\textbf{18}	& \textbf{18}	& \textbf{18} \\				
$football$ \cite{communitystructure1}, $\gamma = 12$		& \textbf{12}	&		\textbf{12}	& 13			& \textbf{12} \\			
$lesmis$ \cite{lesmis}, $\gamma = 10$						& \textbf{10}	&		\textbf{10}	& \textbf{10}	& \textbf{10} \\				
$netscience$ \cite{adjnoun}, $\gamma = 477$			& \textbf{477}	&		\textbf{477}	& 480	& \textbf{477} \\					
$zachary$ \cite{zachary}, $\gamma = 4$					& \textbf{4}	&	\textbf{4} & \textbf{4}	& \textbf{4} \\					
$celegansneural$ \cite{wattsstrogatz}, $\gamma = 16$			& \textbf{16}	&	\textbf{16}		& \textbf{16}	& \textbf{16} \\	
$dolphins$ \cite{dolphins}, $\gamma = 14$					& \textbf{14}	&	\textbf{14}		& \textbf{14}	& \textbf{14} \\					
$astro-ph$ \cite{collaborations}, $\gamma \geq 2928$		& 3138		&		3139	& 3046		& \textbf{2930} \\							
$cond-mat$ \cite{collaborations}, $\gamma \geq 3393$		& 3701		&		3697	& 3501 		& \textbf{3394} \\							
$cond-mat-2003$ \cite{collaborations}, $\gamma \geq 5377$	& 5956			&		5959		& 5601	& \textbf{5379} \\					
$cond-mat-2005$ \cite{collaborations}, $\gamma \geq 6507$		& 7278		&		7254	& 6819		& \textbf{6508} \\						
$hep-th$ \cite{collaborations}, $\gamma \geq 2612$		& 2735	&		2659	& 2653 		& \textbf{2613} \\								
$power$ \cite{wattsstrogatz}, $\gamma \geq 1472$		& 1499 		&		1489	& 1575		& \textbf{1481} \\								
$as-22july06$, $\gamma = 2026$*						& 2223		&		2204	& \textbf{2026}	& \textbf{2026} \\											
$polbooks$, $\gamma = 13$*							& \textbf{13}&	\textbf{13}			& \textbf{13}	& \textbf{13} \\								
\midrule
\multicolumn{5}{l}{DIMACS graphs \cite{johnson::dimacs}}\\\midrule
$anna$, $\gamma = 12$					& \textbf{12}	& \textbf{12}		& \textbf{12}	& \textbf{12} \\							
$homer$, $\gamma = 96$							& \textbf{96}	&		 \textbf{96}	& \textbf{96}	& \textbf{96} \\								
$david$, $\gamma = 2$								& \textbf{2}	&		\textbf{2}	& \textbf{2}	& \textbf{2} \\								
$huck$, $\gamma = 9$							& \textbf{9}	&	\textbf{9}		& \textbf{9}	& \textbf{9} \\\bottomrule		
\end{tabular}
\end{center}
}
\end{table}

RLS$_o$ starts directly with the solution generated by GREEDY, i.e. it both preserves the logarithmic approximation ratio and potentially improves the initial solution. The largest drop in dominating set size compared to GREEDY was from $4817$ to $4585$ obtained by RLS$_o$ for a sample of a network of $50000$ vertices from \textit{Google+}. An improvement from $2761$ to $2714$ was also obtained for a sample of $50000$ vertices from \textit{Pokec}. This indicates that local search process of RLS$_o$ is well suited for finding small dominating sets in very large real-world graphs. It is also worth noting that even though ACO-LS-S provides better results than ACO-LS for this type of graphs, RLS$_o$ performed better than ACO-LS-S for large graphs $gplus\_10000$, $gplus\_20000$, $gplus\_50000$ and $pokec\_50000$. In addition, the results provided by RLS$_o$ were close to the lower bounds, which seems to be encouraging for its use to solve large-scale instances of MDS, as well as its use in hybrid algorithm design. A hybrid algorithm incorporating the ideas of both ACO-based algorithms and RLS$_o$ could be particularly interesting for future algorithm designs.

Next, we performed experiments for graphs studied in \textit{network science}.
These include an adjective-noun adjacency network $adjnoun$ \cite{adjnoun},
American college football league network $football$ \cite{communitystructure1},
coappearance network $lesmis$ for Les Miserables \cite{lesmis},
collaboration network $netscience$ for network science \cite{adjnoun},
network $zachary$ of friendships in a karate club \cite{zachary},
condensed matter collaboration networks $cond-mat$, $cond-mat-2003$ and $cond-mat-2005$ \cite{collaborations},
high energy theory collaboration network $hep-th$ \cite{collaborations},
US power grid network $power$ \cite{wattsstrogatz},
as well as a a snapshot of the Internet on the level of autonomous systems $as-22july06$ and a network $polbooks$ of Krebs' political books. Data also include results for coappearance networks from DIMACS graphs \cite{johnson::dimacs}. These are coappearance networks $anna$ for Anna Karenina, $david$ for David Copperfield, $huck$ for Huckleberry Finn and $homer$ for Iliad and Odyssey.
The obtained results are presented in Table 5 and Table 6, in structures that are identical to Table 3 and Table 4.

Consistently with previous results, the algorithms perform comparably for small graphs. GREEDY produced the best results for 8 networks. ACO-LS, ACO-PP-LS, and ACO-LS-S clearly perform better than GREEDY for small graphs, leading to the best results for 12 networks. Interestingly, ACO-LS-S performs comparably to RLS$_o$ also for the Internet snapshot $as-22july06$. For large collaboration networks $astro-ph$, $cond-mat$, $cond-mat-2003$, $cond-mat-2005$, $hep-th$ and the network $power$, scalability plays a crucial role. RLS$_o$ clearly outperforms the other approaches for these instances, while maintaining its performance for small graphs, leading to the best results obtained for all 19 networks. This was obtained both for the experiments in short runs and long runs.

There are multiple aspects, which make RLS$_o$ successful for large-scale instances of MDS. In addition to the $\mathcal{O}(m)$ complexity of one iteration, where $m$ is the number of edges, the $jump$ operator is able to achieve relatively large changes in the solution structure. Additionally, RLS$_o$ excludes currently redundant vertices directly during the construction using the greedy mapping algorithm. Recalling the structure of Algorithm 1, the formulation of step 7 ensures that a vertex is added to the dominating set only if it is non-dominated or it has at least one non-dominated neighbour. This way, RLS$_o$ effectively avoids the need for a consequent elimination of redundant vertices, while keeping the construction process efficient for sparse graphs.

\subsection{Results of MSRLS$_o$ in MWDS}

However, RLS$_o$ is suitable only for MDS by its design, as it prefers dominating sets with lower cardinality. As MWDS is a problem motivated by applications in wireless networks, we have tested the multi-start algorithm MSRLS$_o$ for instances derived from these applications.  We have used SMPI instances for which results of various algorithms are widely reported in the literature. These instances are divided into categories T1 and T2 and each instance group consists of $10$ instances with the same number of vertices and edges. This allowed us to compare MSRLS$_o$ to a wider range of experimental approaches to solve MWDS. It is however worth mentioning that the following results only represent a comparison for this type of instances. Our experimental experience from this study indicates that one unfortunately cannot easily deduce how well state-of-the-art algorithms for MWDS perform in MDS for large-scale social and information networks and vice versa. This is likely influenced by different sizes and structural properties of instances derived from real-world applications of MDS and MWDS.

MSRLS$_o$ was used with a maximum of $c_{max} = 5000$ search cycles, with each cycle cut off after $i_{max} = 2000$ iterations by default and $i_{best} = 100000$ iterations in case that the current cycle has led to an improvement of the best solution found so far. The time limit was $60$ minutes for each instance but the search was much quicker in a vast majority of cases. However, MSRLS$_o$ appears to be slower than other algorithms for MWDS, suggesting that its hybridisation with more specialised ideas for MWDS could be the way forward.

The results for T1 and T2 instances are given in Table 7 and Table 8, respectively. MSRLS$_o$ has been compared to algorithms Raka-ACO \cite{jovanovic2010ant}, ACO-LS and ACO-PP-LS \cite{dominatingweight}, EA/G-IR \cite{chaurasia2015hybrid}, HMA \cite{lin2016effective}, Hyb-R-PBIG \cite{bouamama2016hybrid} and CC$^{2}$FS \cite{wang2017local}. The results reported for MSRLS$_o$ are the average values obtained for each instance group over $5$ independent runs for each instance.

These results show that MSRLS$_o$ is able to provide relatively good results for MWDS. Its performance is comparable to EA/G-IR and HMA, which are quite specific hybrid evolutionary algorithms for MWDS. MSRLS$_o$ uses a set of quite general ideas, originally designed to solve MDS in large-scale complex networks. Hyb-R-PBIG and CC$^{2}$FS generally perform better, even though these algorithms have very specific designs for MWDS. These results indicate that MSRLS$_o$ could be beneficial specifically as a subroutine within future hybrid algorithms for MWDS. It would also be interesting to further use and adapt the algorithms for MWDS to solve MDS for large-scale social and information networks.

\begin{table}
\caption{Computational results of MSRLS$_o$ in comparison to the state-of-the-art algorithms for MWDS for T1 instances.}
{\tiny
\begin{center}
\begin{tabular}{l l l l l l l l l}\toprule
instance								& MSRLS$_o$ 	& Raka-			& ACO-LS		&	ACO-			& EA/G-IR		& HMA					&	Hyb-					& CC$^{2}$FS	\\
group										&							& ACO				& 					&	PP-LS			& 					& 						&	R-PBIG			& 						\\
\midrule
$50\_50$        & 531.3              & 539.8              &       531.3   &       532.6   &       532.9   &       531.8   &       531.3   &       531.3   \\
$50\_100$        & 370.9              &  391.9             &       371.2   &       371.5   &       371.5   &       371.2   &       370.9   &       370.9   \\
$50\_250$        & 175.7	              &  195.3             &       176     &       175.7   &       175.7   &       176.4   &       175.7   &       175.7   \\
$50\_500$        & 94.9              &   112.8            &       94.9    &       95.2    &       94.9    &       96.2    &       94.9    &       94.9    \\
$50\_750$        & 63.1              &   69            &       63.1    &       63.2    &       63.3    &       63.3    &       63.1    &       63.3    \\
$50\_1000$        & 43.2              &   44.7            &       41.5    &       41.5    &       41.5    &       41.5    &       41.5    &       41.5    \\
$100\_100$        & 1061              &   1087.2            &       1066.9  &       1065.4  &       1065.5  &       1064.9  &       1061    &       1061    \\
$100\_250$        & 618.9              &    698.7           &       627.2   &       627.4   &       620     &       623.1   &       618.9   &       618.9   \\
$100\_500$        & 355.6               &    442.8           &       362.5   &       363.2   &       355.9   &       356.8   &       355.6   &       355.6   \\
$100\_750$        & 255.8              &    313.7           &       263.5   &       265     &       256.7   &       258.4   &       255.8   &       255.8   \\
$100\_1000$        & 203.6              &    247.8           &       209.2   &       208.8   &       203.6   &       205.9   &       203.6   &       203.6   \\
$100\_2000$        & 107.4              &    125.9           &       108.1   &       108.4   &       108.1   &       107.8   &       107.4   &       107.4   \\
$150\_150$        & 1582.1              &    1630.1           &       1582.8  &       1585.2  &       1587.4  &       1585.3  &       1580.5  &       1580.5  \\
$150\_250$        & 1221.7              &    1317.7           &       1237.2  &       1238.3  &       1224.5  &       1231.8  &       1218.2  &       1218.2  \\
$150\_500$        & 746.9              &  899.9             &       767.7   &       768.6   &       755.3   &       749.5   &       744.6   &       744.6   \\
$150\_750$        & 549.1              &   674.4            &       565     &       562.8   &       550.8   &       550.2   &       546.8   &       546.1   \\
$150\_1000$        & 434.9              &  540.7             &       446.8   &       448.3   &       435.2   &       435.7   &       433.1   &       432.9   \\
$150\_2000$        & 241.1              &   293.1            &       259.4   &       255.6   &       241.5   &       244.2   &       241.8   &       240.8   \\
$150\_3000$        & 166.9              &    204.7           &       173.4   &       175.2   &       168.1   &       168.4   &       167.8   &       166.9   \\
$200\_250$        & 1917.9              &   2039.2            &       1934.3  &       1927    &       1924.1  &       1912.1  &       1909.7  &       1910.4  \\
$200\_500$        & 1242.1              &   1389.4            &       1259.7  &       1260.8  &       1251.3  &       1245.7  &       1234    &       1232.8  \\
$200\_750$        & 923.1              &    1096.2           &       938.7   &       940.1   &       927.3   &       926.1   &       913.8   &       911.2   \\
$200\_1000$        & 737.7              &    869.9           &       751.2   &       753.7   &       731.1   &       727.4   &       726 &   724     \\
$200\_2000$        & 423.1              &     524.1          &       440.2   &       444.7   &       417     &       421.2   &       414.7   &       412.7   \\
$200\_3000$        & 298.4              &     385.7          &       309.9   &       315.2   &       294.7   &       297.9   &       296 &   292.8   \\
$250\_250$        & 2646.7              &  -             &       2655.4  &       2655.4  &       2653.7  &       2653.4  &       2633    &       2633.4  \\
$250\_500$        & 1827.1              &  -             &       1850.3  &       1847.9  &       1853.3  &       1828.5  &       1806.1  &       1805.9  \\
$250\_750$        & 1392.9               & -              &       1405.2  &       1405.5  &       1399.2  &       1389.4  &       1366.9  &       1362.2  \\
$250\_1000$        & 1119.6              & -              &       1127.1  &       1122.9  &       1114.9  &       1109.5  &       1092.8  &       1091.1  \\
$250\_2000$        & 651.5              & -              &       672.8   &       676.4   &       637.5   &       635.3   &       624.2   &       621.9   \\
$250\_3000$        & 466.4              & -              &       474.1   &       476.3   &       456.3   &       456.6   &       452.5   &       447.9   \\
$250\_5000$        & 297.8              & -              &       310.4   &       308.7   &       291.8   &       292.8   &       293.1   &       289.5   \\
\midrule
$300\_300$        & 3202.4              & -              &       3198.5  &       3205.9  &       3213.7  &       3199.3  &       3175.4  &       3178.6\\
$300\_500$        & 2468.6              & -              &       2479.2  &       2473.3  &       2474.8  &       2464.4  &       2435.6  &       2438.1\\
$300\_750$        & 1897.1              & -              &       1903.3  &       1913.9  &       1896.3  &       1884.6  &       1856.8  &       1854.6\\
$300\_1000$        & 1539.2              & -              &       1552.5  &       1555.8  &       1531    &       1518.4  &       1498.6  &       1495\\
$300\_2000$        & 901.3               & -              &       916.8   &       916.5   &       880.1   &       878.7   &       870.1   &       862.5\\
$300\_3000$        & 663.5              & -              &       667.8   &       670.7   &       638.2   &       640.9   &       628.5   &       624.3\\
$300\_5000$        & 428.5              & -              &       437.4   &       435.9   &       415.7   &       411.7   &       410 &   406.1\\
$500\_500$        & 5389.2              &   5476.3            &       5398.3  &       5387.7  &       5380.1  &       5392.1  &       5304.7  &       5305.7\\
$500\_1000$        & 3716.3              &   4069.8            &       3714.8  &       3698.3  &       3695.2  &       3678.3  &       3607.3  &       3607.8\\
$500\_2000$        & 2291.2              &   2627.5            &       2277.6  &       2275.9  &       2264.3  &       2223.7  &       2197.2  &       2181\\
$500\_5000$        & 1138.1              &    1398.5           &       1115.3  &       1110.2  &       1083.5  &       1074.2  &       1052.1  &       1043.3\\
$500\_10000$        & 634.5              &     825.7          &       652.8   &       650.9   &       606.8   &       595.4   &       597.5   &       587.2\\
$800\_1000$        & 7833.7             &      8098.9         &       8117.6  &       8068    &       7792.2  &       7839.9  &       7655    &       7663.4\\
$800\_2000$        &  5224.6             &       5739.9        &       5389.9  &       5389.6  &       5160.7  &       5100.7  &       5002.8  &       4982.1\\
$800\_5000$        & 2626.9              &     3116.5          &       2616    &       2607.9  &       2561.9  &       2495.7  &       2469.2  &       2441.2\\
$800\_10000$        & 1526.2              &     1923          &       1525.7  &       1535.3  &       1497    &       1459.8  &       1414.8  &       1395.6\\
$1000\_1000$        & 10766.7      &      10924.4         &       11035.5 &       11022.9 &       10771.7 &       10863.3 &       10574.4 &       10585.3\\
$1000\_5000$        & 3947          &     4662.7          &       4012    &       4029.8  &       3876.3  &       3742.8  &       3699.7  &       3671.8\\
$1000\_10000$        & 2283.2          &    2890.3           &       2314.9  &       2306.6  &       2265.1  &       2193.7  &       2138.1  &       2109\\
\bottomrule
\end{tabular}
\end{center}
}
\end{table}

\begin{table}
\caption{Computational results of MSRLS$_o$ in comparison to the state-of-the-art algorithms for MWDS for T2 instances.}
{\tiny
\begin{center}
\begin{tabular}{l l l l l l l l l}\toprule
instance									& MSRLS$_o$ & Raka-			& ACO-LS		&	ACO-			& EA/G-IR		& HMA					&	Hyb-					& CC$^{2}$FS	\\
group											&						& ACO				& 					&	PP-LS			& 					& 						&	R-PBIG			& 						\\
\midrule
$50\_50$        &  60.8             &    62.3           &       60.8    &       60.8    &       60.8    &       60.8    &       60.8    &       60.8    \\
$50\_100$        &  90.3             &    98.4           &       90.3    &       90.3    &       90.3    &       90.3    &       90.3    &       90.3    \\
$50\_250$        &  146.7             &   202.4            &       146.7   &       146.7   &       146.7   &       146.7   &       146.7   &       146.7   \\
$50\_500$        & 179.9              &    312.9           &       179.9   &       179.9   &       179.9   &       179.9   &       179.9   &       179.9   \\
$50\_750$        & 171.1              &    386.3           &       171.1   &       171.1   &       171.1   &       171.1   &       171.1   &       171.1   \\
$50\_1000$        & 146.5              &  -             &       146.5   &       146.5   &       146.5   &       146.5   &       146.5   &       146.5   \\
$100\_100$        &  123.5             &    126.5           &       123.6   &       123.5   &       123.5   &       124.4   &       123.5   &       123.5   \\
$100\_250$        & 210.8              &     236.6          &       210.2   &       210.4   &       209.2   &       210.4   &       209.2   &       209.2   \\
$100\_500$        & 305.9              &      404.8         &       307.8   &       308.4   &       305.7   &       307.1   &       305.7   &       305.7   \\
$100\_750$        & 384.5              &      615.1         &       385.7   &       386.3   &       384.5   &       384.8   &       384.5   &       384.5   \\
$100\_1000$        & 427.3              &      697.3         &       430.3   &       430.3   &       427.3   &       428     &       427.3   &       427.3   \\
$100\_2000$        & 551              &       1193.9        &       558.8   &       559.8   &       550.6   &       552.2   &       550.6   &       550.6   \\
$150\_150$        & 184.5              &     190.1          &       184.7   &       184.9   &       184.5   &       185.8   &       184.5   &       184.5   \\
$150\_250$        & 234              &     253.9          &       233.2   &       233.4   &       232.8   &       234.1   &       232.8   &       232.8   \\
$150\_500$        & 353.9              &    443.2           &       351.9   &       351.9   &       349.7   &       350.9   &       349.5   &       349.5   \\
$150\_750$        & 458.9              &    623.3           &       456.9   &       454.7   &       452.4   &       435.5   &       452.4   &       452.4   \\
$150\_1000$        & 549.1              &     825.3          &       551.4   &       549     &       548.2   &       549.4   &       547.2   &       547.2   \\
$150\_2000$        & 720.1              &      1436.4         &       725.7   &       725.7   &       720.1   &       722.2   &       720.1   &       720.1   \\
$150\_3000$        & 792.6              &      1751.9         &       794     &       806.2   &       792.4   &       797.8   &       792.4   &       792.4   \\
$200\_250$        & 273.2              &      293.2         &       272.6   &       272.6   &       272.3   &       273.8   &       271.7   &       271.7   \\
$200\_500$        & 401.8              &       456.5        &       388.6   &       388.4   &       388.4   &       389.1   &       386.7   &       386.7   \\
$200\_750$        & 503.4              &       657.9        &       501.7   &       501.4   &       497.2   &       500.3   &       497.1   &       497.1   \\
$200\_1000$        & 609.7              &     829.2          &       605.9   &       605.8   &       598.2   &       606.6   &       596.8   &       596.8   \\
$200\_2000$        & 890.2              &    1626           &       891     &       892.9   &       885.8   &       890.3   &       884.6   &       884.6   \\
$200\_3000$        &  1029.2             &    2210.3           &       1027    &       1034.4  &       1019.7  &       1026.2  &       1019.2  &       1019.2  \\
$250\_250$        & 306.6              &   -            &       306.5   &       306.7   &       306.5   &       310.4   &       306 &   306.1   \\
$250\_500$        & 454.3              &  -             &       443.8   &       443.2   &       441.6   &       445.8   &       440.7   &       440.7   \\
$250\_750$        & 583.8              &  -             &       573.1   &       575.9   &       569.2   &       573.1   &       567.4   &       567.4   \\
$250\_1000$        & 689.6              &  -             &       671.8   &       675.1   &       671.7   &       676.7   &       668.6   &       668.6   \\
$250\_2000$        & 1044.7              & -              &       1033.9  &       1031.5  &       1010.3  &       1025.8  &       1007    &       1007    \\
$250\_3000$        & 1262.4              &  -             &       1288.5  &       1277    &       1250.6  &       1261.4  &       1250.6  &       1250.6  \\
$250\_5000$        & 1472.9              &  -             &       1493.6  &       1520.1  &       1464.2  &       1484    &       1464.2  &       1464.2  \\
\midrule
$300\_300$        & 371.2              &    -           &       371.1   &       371.1   &       370.5   &       373.9   &       369.9   &       369.9\\
$300\_500$        & 485.8              &   -           &       480.8   &       481.2   &       480     &       484     &       477.8   &       477.8\\
$300\_750$        & 634.9               &   -            &       621.6   &       618.3   &       613.8   &       620.1   &       613.3   &       613.3\\
$300\_1000$        & 755.8              &   -            &       744.9   &       743.5   &       742.2   &       745.1   &       737.7   &       737.9\\
$300\_2000$        & 1118              &   -            &       1111.6  &       1107.5  &       1094.9  &       1110.3   &       1093.8  &       1093.8\\
$300\_3000$        & 1392.5              &   -            &       1422.8  &       1415.3  &       1359.5  &       1378.9  &       1358.5  &       1358.5\\
$300\_5000$        & 1701.3              &  -             &       1712.1  &       1698.6  &       1683.6  &       1692.6  &       1682.7  &       1682.7\\
$500\_500$        & 627              &    651.2           &       627.5   &       627.3   &       625.8   &       633.4   &       623.6   &       623.6\\
$500\_1000$        & 934.9              &   1018.1            &       913     &       912.6   &       906     &       912     &       899.6   &       899.8\\
$500\_2000$        & 1408.6              &    1871.8           &       1384.9  &       1383.9  &       1376.7  &       1394.1  &       1362.2  &       1363.3\\
$500\_5000$        & 2401.7              &   4299.8            &       2459.1  &       2468.8  &       2340.3  &       2388.3  &       2326.6  &       2333.7\\
$500\_10000$        & 3261.5              &   8543.5            &       3377.9  &       3369.4  &       3216.4  &       3259.6  &       3211.5  &       3211.5\\
$800\_1000$        & 1126.2              &     1171.2          &       1126.4  &       1125.1  &       1107.9  &       1131.3  &       1103.9  &       1104.3\\
$800\_2000$        & 1709.9               &     1938.7          &       1693.7  &       1697.9  &       1641.7  &       1681.9  &       1630.8  &       1632.3\\
$800\_5000$        & 2990.4             &     4439          &       3121.9  &       3120.9  &       2939.3  &       2963.8  &       2876.1  &       2878.5\\
$800\_1000$        & 4272.5              &     8951.1          &       4404.1  &       4447.9  &       4155.1  &       4226.6  &       4103.3  &       4105.6\\
$1000\_1000$        & 1247.6              &      1289.3         &       1259.3  &       1258.6  &       1240.8  &       1270.9  &       1237.5  &       1237.7\\
$1000\_5000$        & 3340.5              &    4720.1           &       3411.6  &       3415.1  &       3222    &       3317.6  &       3172.9  &       3178.7\\
$1000\_10000$        & 4935.5              &    9407.7           &       5129.1  &       5101.9  &       4798.6  &       4937.9  &       4704.8  &       4711.8\\
\bottomrule
\end{tabular}
\end{center}
}
\end{table}

\subsection{Application of RLS$_o$ in Graph Mining and Discussion}

In the previous investigations, we demonstrated that RLS$_o$ is a suitable algorithm for finding small dominating sets in large graphs, including real-world complex networks. In order to study these results in a more applied context, we now discuss the application of small dominating sets in graph mining.

\textit{Graph mining} is an area focused on study of large-scale real-world graphs, their typical structure and design of algorithms for knowledge discovery in these graphs \cite{graphmining}. \textit{Graph clustering and community detection} problems are among the most widely studied topics in this area, as well as in contemporary computer science \cite{communitystructure1,wirelesscommunity,leskovec2,gacommunities,graphclustering}. On the other hand, there is a large spectrum of possible definitions of clusters, communities and evaluation criteria for quality of these structures \cite{leskovec2,graphclustering}. Therefore, we briefly present a more unified view on these methods and their similarity to this research.

\begin{figure}
\begin{center}
\includegraphics[scale=0.85, bb=0 0 189 189]{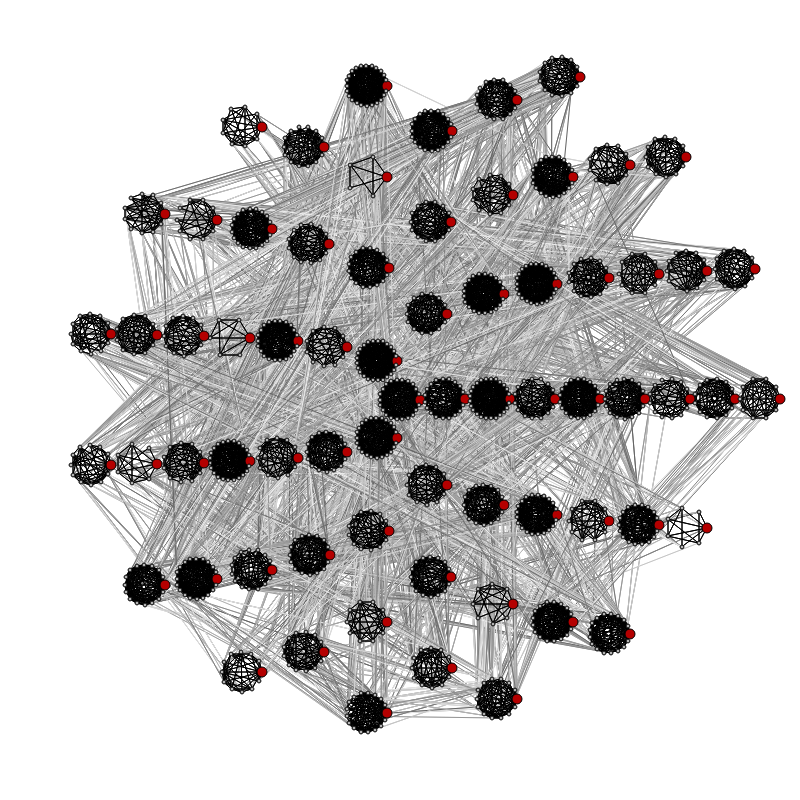}
\includegraphics[scale=0.85, bb=0 0 189 189]{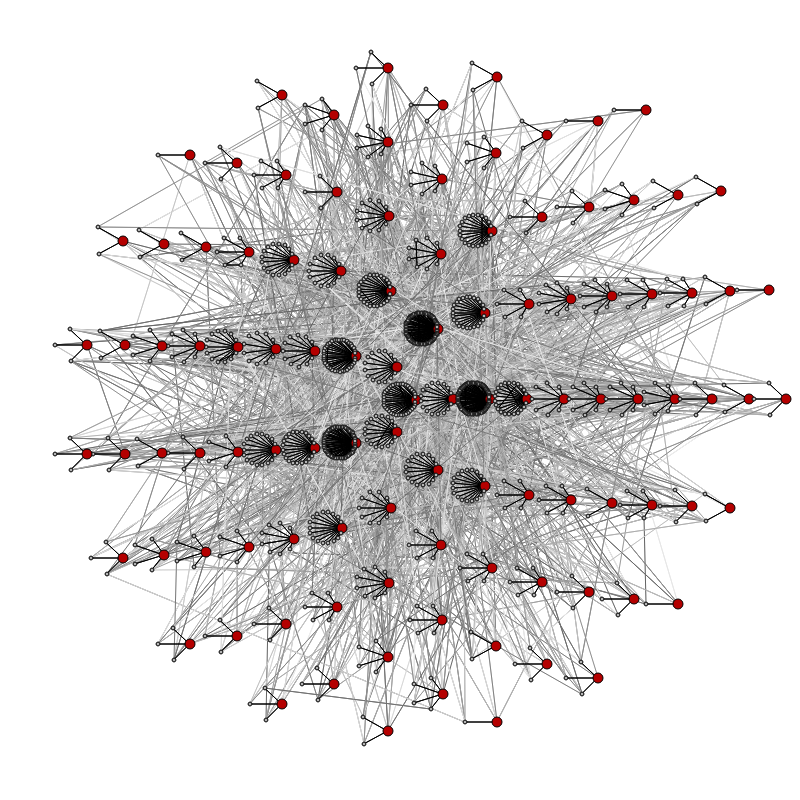}
\end{center}
\caption{Visualisations of dominating sets obtained by RLS$_o$ for a unit disk graph on $1000$ vertices with $range = 150$ and grid size $2000 \times 2000$ (on the left-hand side) and for a scale-free network generated by BA model on $1000$ vertices with the number of incoming edges $w = 4$ (on the right hand side). Vertices in the dominating set are marked red and their neighbours are grouped into clusters. Dominating set of the unit disk graph consists of $68$ vertices and leads to relatively larger and denser clusters, while on the right hand side, we have the dominating set with $104$ vertices for the scale-free network, revealing a pseudo-star structure of the network, containing both larger and smaller clusters.}
\end{figure}

One of the most classical approaches to model network communities is the \textit{k-medoids problem} \cite{simplefastkmedoids}, in which one searches for $k$ medoids such that partitioning the vertices around the closest medoids gives minimum possible distance within the clusters. While in k-medoids, one aims to minimise the distance while number of clusters is fixed, dominating set tackles the problem by minimising the number of clusters while distance to a vertex of the dominating set is at most $1$. This formulation reminds one of the small world properties of many real-world networks \cite{wattsstrogatz}.

On the other hand, dominating set algorithms have also found their applications in clustering of real-world networked data. This includes clustering of mobile ad-hoc networks \cite{indukuri2011dominating}, as well as clustering algorithms for small satellite networks \cite{qin2012weight}. Social, information or biological network clustering represents another interesting perspective.

Figure 2 depicts the dominating sets found by RLS$_o$ and the corresponding clusters in dominance drawings for a unit disk graph (on the left-hand side) and a scale-free network (on the right-hand side). The unit disk graph was generated for $1000$ vertices with $range = 150$ and grid size $2000 \times 2000$. RLS$_o$ was able to find a dominating set with $68$ vertices, while ACO-LS found a dominating set with $71$ vertices. The right-hand side of Figure 2 presents a similar result for scale-free network generated by BA model on $1000$ vertices with $w = 4$ incoming edges per vertex. RLS$_o$ found a dominating set of size $104$, while ACO-LS used $107$ vertices to form a dominating set. Both for unit disk graphs and scale-free networks, RLS$_o$ leads to a slightly more compact representation of the network than alternative approaches.

\begin{figure}
\begin{center}
\includegraphics[scale=0.85, bb=0 0 189 189]{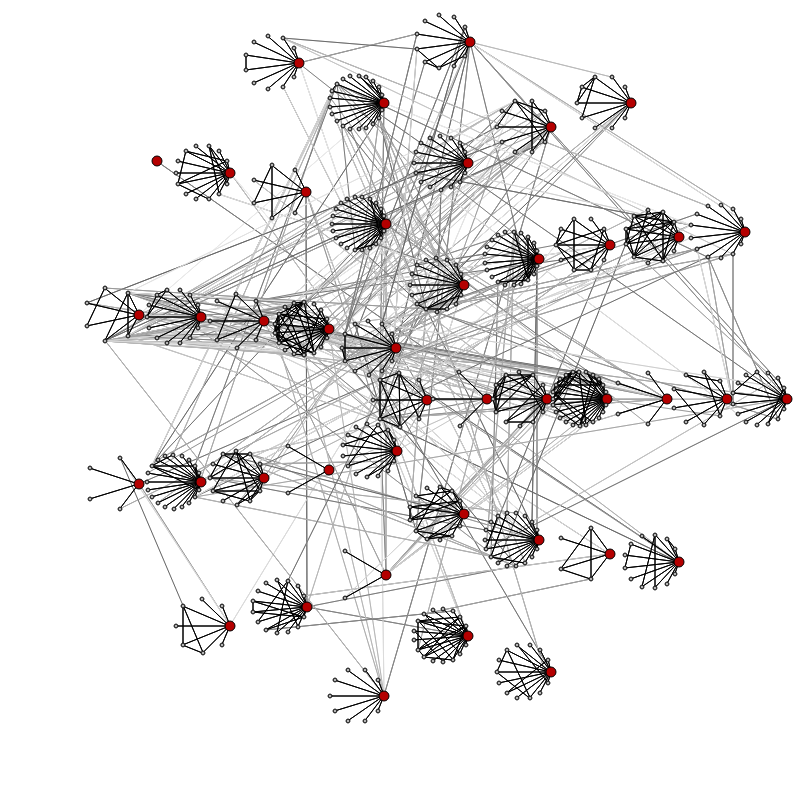}
\includegraphics[scale=0.85, bb=0 0 189 189]{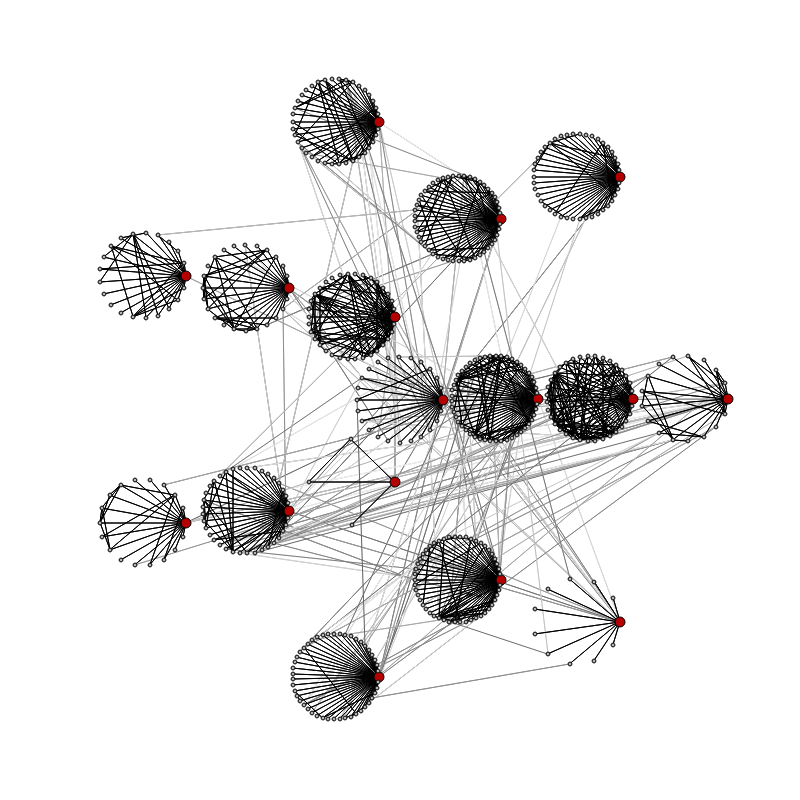}
\end{center}
\caption{Visualisations of smallest dominating sets obtained for samples of social networks \textit{Google+} and \textit{Pokec} with $500$ vertices. On the left hand size, we visualise a dominating set with $42$ vertices for \textit{Google+} and on the right hand size, we have a dominating set with $16$ vertices for \textit{Pokec}.}
\end{figure}

In Figure 3, we illustrate the dominating sets for samples from \textit{Google+} and \textit{Pokec} with $500$ vertices. The sample from \textit{Google+} is sparser, since it is created from publicly available connections. For both graphs, dominating sets reveal the community structure reliably. These communities are tightly partitioned around the dominating set vertices, with each community member being adjacent to the corresponding member of the dominating set.

These results indicate that small dominating sets might be suitable as a representation of community structure for different types of graphs. If the network is a small world, dominating set will generally be a solid choice for the clustering problem. In addition, if the network is large, RLS$_o$ will be a good choice of a scalable heuristic to compute a small dominating set.

\section{Conclusions}

We proposed an \textit{order-based randomised local search} (RLS$_o$) algorithm for minimum dominating set problem. Evaluation of RLS$_o$ was carried out for \textit{unit disk graphs}, \textit{scale-free networks} generated by Barab\'{a}si-Albert model, as well as real-world graphs including \textit{samples of two social networks} and graphs studied in the field of \textit{network science}. A multi-start variant MSRLS$_o$ of the algorithm was also proposed for the minimum weight dominating set (MWDS) problem.

Experimental results indicate that RLS$_o$ performs better than a classical greedy approximation algorithm for the problem, as well as hybrid heuristics based on ant colony optimisation and local search (ACO-LS) and ant colony optimisation with preprocessing and local search (ACO-PP-LS). The results obtained by RLS$_o$ indicate that it is a suitable approach to solve large-scale instances of the problem. As another result, we also designed a simple modification of ACO-LS, which allows transitions only between adjacent vertices during solution construction, making it better scalable to solve the problem in large-scale social and information networks. However, RLS$_o$ is the algorithm, which produced the best results in our experimental studies.

In this context, interesting open problems include theoretical analysis of behaviour of RLS$_o$ and the ant-based algorithms. Although one cannot expect better than logarithmic approximation ratio for scale-free networks \cite{powerlawinapprox}, interesting results might still be obtainable for special cases of graphs, which are interesting for real-world applications. Additionally, tighter lower bounds for minimum dominating set size in large real-world graphs should be of a high interest.

Last but not least, illustration of the application of our approach in graph mining opens its possible further use in application areas. Partitioning of the network around dominating set vertices leads to clusters, for which it holds that every vertex is in distance to a vertex of the dominating set, which is at most $1$. Therefore, RLS$_o$ should be interesting for applications, which require a fast and highly scalable technique and for which the corresponding network exhibits small-world properties.

\bibliography{common}{}
\bibliographystyle{plain}

\end{document}